\documentclass[prb,amsmath,twocolumn]{revtex4-2}
\usepackage[utf8]{inputenc}

\usepackage{amssymb}
\usepackage{amsmath}
\usepackage{amsfonts}
\usepackage{bbold}
\usepackage{bm}
\usepackage{bbm}
\usepackage{braket}
\usepackage{comment}
\usepackage{color}
\usepackage{epsfig}
\usepackage{graphicx}
\usepackage{hyperref}
\usepackage{listings}
\usepackage{soul}
\usepackage{tikz}
\usepackage{multirow}
\usepackage[linesnumbered,ruled]{algorithm2e}

\begin{document}

\title{Localisation and Sub-Diffusive Transport in Quantum Spin Chains With Dilute Disorder}
\author{S.~J.~Thomson}
\affiliation{Dahlem Centre for Complex Quantum Systems, Freie Universität, 14195 Berlin, Germany}
\date{\today}

\begin{abstract}
    It is widely believed that many-body localisation in one dimension is fragile and can be easily destroyed by thermal inclusions, however there are still many open questions regarding the stability of the localised phase and under what conditions it breaks down. Here I construct models with dilute disorder, which interpolate between translationally invariant and fully random models, in order to study the breakdown of localisation. This opens up the possibility to controllably increase the density of thermal regions and examine the breakdown of localisation as this density is increased. At strong disorder, the numerical results are consistent with commonly-used diagnostics for localisation even when the concentration of thermalising regions is high. At moderate disorder, I present evidence for slow dynamics and sub-diffusive transport across a large region of the phase diagram, suggestive of a `bad metal' phase. This suggests that dilute disorder may be a useful effective model for studying Griffiths effects in many-body localisation, and perhaps also in a wider class of disordered systems.
\end{abstract}

\maketitle

\section{Introduction}

The study of many-body quantum systems which fail to thermalise is a major frontier in modern condensed matter physics, and by now there are many examples of scenarios where thermalisation may be avoided, most commonly through the addition of some form of disorder~\cite{Anderson58,Fleishman+80,Basko+06,Huse+13,HuseNandkishoreOganesyanPRB14,Imbrie16a,Imbrie+17,Alet+18,AbaninEtAlRMP19}, but also even in a variety of disorder-free systems~\cite{Smith+17,Brenes+18,Schulz+19,vanNieuwenburg+19,Karpov+21,Halimeh+22a,Halimeh+22b,Lang+22,Chakraborty+22}. Similar effects can even be achieved even by preparing certain systems in finely-tuned initial states without any form of disorder, as in the case of quantum many-body scars~\cite{Turner+18}.

The focus of this work is many-body localisation~\cite{Huse+13,HuseNandkishoreOganesyanPRB14,Imbrie16a,Imbrie+17,Alet+18,AbaninEtAlRMP19}, the interacting variant of Anderson localisation~\cite{Anderson58,Fleishman+80} where an isolated many-body quantum system can become localised at all energy scales via the addition of a random on-site chemical potential or magnetic field~\cite{Basko+06}. Recent years have seen huge progress in understanding many-body localisation -- particularly from the point of view of local integrals of motion~\cite{Ros+15,Rademaker+16,Rademaker+17,Imbrie+17,Thomson+18,Thomson+20b,Thomson+21} -- as well as in establishing under what conditions it can and can not exist. Remarkably, there is even a class of spin chains for which an analytical proof of many-body localisation exists~\cite{Imbrie16a}, subject to the assumption of limited level attraction, which is widely considered to be a reasonable assumption. Behaviour consistent with many-body localisation has been experimentally observed in a variety of one- and even two-dimensional systems~\cite{Schreiber+15,Bordia+16,Luschen+17,Bordia+17,Smith+16,Choi+16,Sbroscia+20}, though its stability remains under intense discussion. In particular, it is widely believed that many-body localisation is stable only in one-dimensional systems with short-range couplings, as otherwise localisation has been argued to be unstable due to what is now known as the avalanche effect, where rare regions of anomalously low disorder can form ergodic bubbles which can grow and eventually cause the entire system to thermalise~\cite{Thiery+18}.

Recently, however, the stability of many-body localisation even in one dimension has come into question, with some works suggesting that the localisation transition takes place at a much higher disorder strength than previously expected~\cite{Doggen+18}, and others suggesting that there may be no transition at all in the thermodynamic limit~\cite{Suntajs+20a,Suntajs+20b,Sels+21a,Sels+21b,Sels22}. It is therefore of great interest to study in more detail the breakdown of many-body localisation and characterise just how robust this phase really is. Whether or not it constitutes a stable phase in the thermodynamic limit, it is clear that upon increasing the disorder strength the relaxation dynamics of the system enters a regime that either does not thermalise at all, or does so only on extremely long times. In the following I shall refer to this regime as `many-body localised', and to the onset of these slow dynamics as the `many-body localisation transition'. In particular, there has been a great deal of work studying the transport properties close to the many-body localisation transition, with a large body of evidence pointing towards the presence of a Griffths-like subdiffusive regime and other anomalous transport properties both in microscopic models of many-body systems~\cite{BarLev+15,Khait+16,DeRoeck+20,Luitz+16,LuitzEtALPRB16,Varma+17,Znidaric+16} and proxy models which investigate MBL via Anderson localisation in Fock space~\cite{DeTomasi+20,Prelovsek+17,Bera+17,Schiro+19,Biroli+17,Bera+18,Khaymovich+21,Colmenarez+22}. Gaining further insights these anomalous transport properties could lead to an improved understanding of the many-body localisation transition itself.

In this work I take a new approach to studying the breakdown of the many-body localised phase. I construct a model in which homogeneous thermalising regions can be controllably added, and ask what happens to many-body localisation as the density of these thermalising regions is varied. Previous work~\cite{Gopalakrishnan+15,Thiery+18} has suggested that rare thermalising regions should play a key role close to the transition. These models allow us to investigate the effects of both rare thermal regions and rare impurities, which in one dimension can form large bottlenecks that can have a significant effect on the transport properties of the system, even if the majority of lattice sites are in locally ergodic regions. This form of disorder may be a useful toy model for Griffiths effects~\cite{Vojta10} and rare region physics that are believed to play an important role in the many-body localisation transition in systems with conventional random disorder~\cite{Agrawal+15,Schiro+19}, as well as in other paradigmatic examples of disordered phases of matter such as the Bose glass~\cite{Fisher+89}. 

\begin{figure*}
    \centering
    \includegraphics[width=\linewidth]{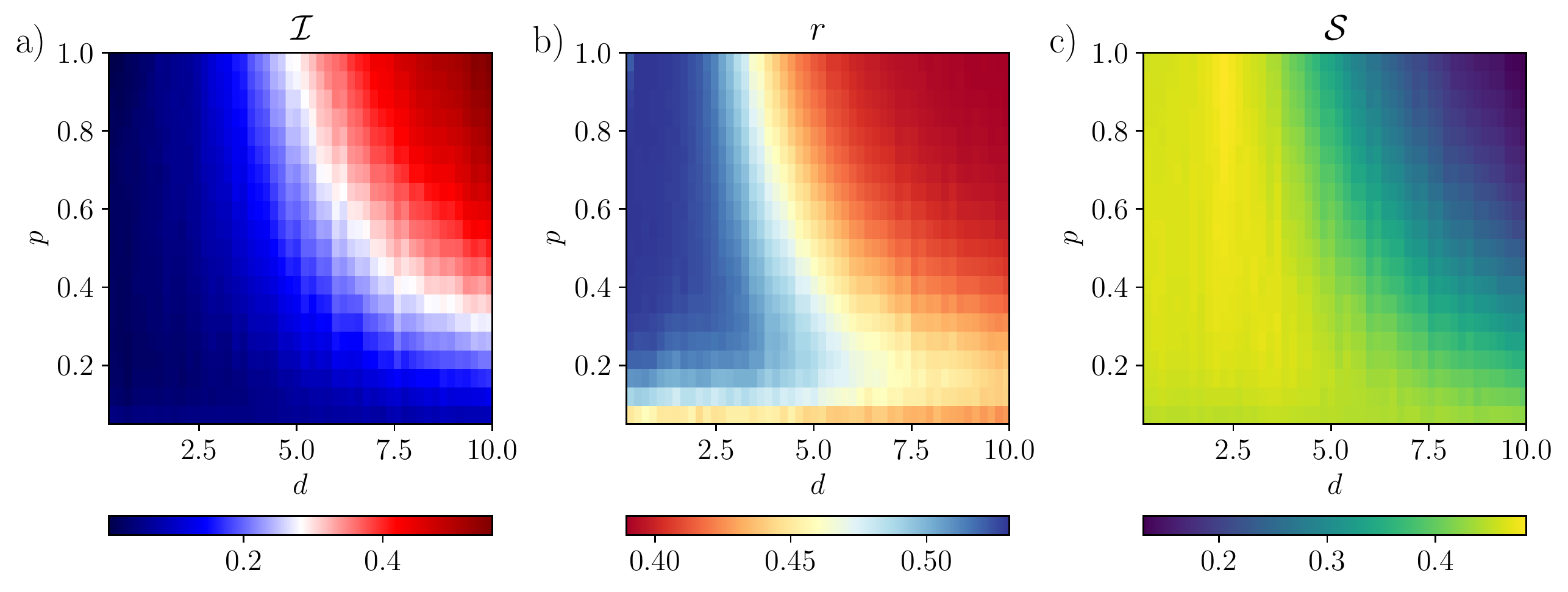}
    \caption{Numerical results for the case of dilute random disorder, for a system size $L=14$ and averaged over $N_s=512$ disorder realisations. a) The imbalance $\mathcal{I}$ long after a quench ($tJ_0=10^3$) from a Néel state. b) The averaged level spacing statistics $r$, which vary from $r \sim 0.39$ in a localised phase to $r \sim 0.53$ in an ergodic phase. c) The long-time behaviour of the entanglement entropy $\mathcal{S}$, again at a time $tJ_0=10^3$ following a quench. In all plots, the upper left corner is chaotic, while the upper right corner is localised. The lower left corner is chaotic, except for close to $p=0$ where integrability is approximately restored. The lower right corner exhibits a small persistent imbalance, level statistics that are close to Poisson, and an entanglement entropy that grows slower than expected in a chaotic system. All of these are suggestive of a `bad metal' regime existing over an extended region of the phase diagram.}
    \label{fig.dilute_random}
\end{figure*}

\section{Model}

I will focus on the XXZ model of strongly interacting spins in one dimension with nearest-neighbour interactions, subject to two different choices of on-site potential. The Hamiltonian for a chain of length $L$ with open boundary conditions is given by:
\begin{align}
H = \sum_i h_i S_i^z + J_0 \sum_{i} (S^{x}_i S^{x}_{i+1} + S^{y}_i S^{y}_{i+1}) + J_z  \sum_{i} S^z_i S^z_{i+1}.
\end{align}
In the following, all Hamiltonian parameters will be measured in units of $J_0=1$, and I will set $J_z=J_0$. All calculations will be performed in the zero magnetisation sector.
I shall consider the on-site terms $h_i$ drawn from two different distributions.
The first case, which I shall call \emph{dilute random disorder}, is described by a potential which is randomly chosen to be either equal to a uniform value or drawn randomly from a box distribution of width $2d$:
\begin{align}
h _i =
    \begin{cases}
      1, & \text{with probability}\ (1-p), \\
      \in [-d,d], & \text{with probability}\ p.
    \end{cases}
\end{align}
In the limit $p \to 1$, this model reduces to the `standard model' of MBL, namely a fully random system with a completely disordered on-site potential, which has been widely studied in the literature and its properties are by now well-established~\cite{Pal+10,Bardarson+12,Nanduri+14,Luitz+15}. In the opposite limit of $p \to 0$, the model instead becomes a translationally invariant integrable system, the properties of which are also well-established. By varying $p$, it is possible to interpolate between these two limits and examine at which point localisation breaks down. For a given value of $p$, the chance of a region of length $L$ which is completely homogeneous decays exponentially with the length of the region, and is given by $(1-p)^L$. Conversely, the chance to find a region of length $L$ which is entirely random is $p^L$, and for $p=0.5$ at fixed $L$ both types of region are equally likely to occur. The expected size of the largest disordered region is given by $R_{L}(p) \approx \log_{1/p}(L(1-p)) = \log(L(1-p))/\log(1/p)$, however due to the skewness of the distribution, the largest disordered region can in fact be much larger than this expectation might suggest~\cite{Schilling90}. Further details on the distribution of rare regions are given in Appendix~\ref{app.dist}. It is important to note that by the nature of random systems, it is possible -- indeed, inevitable -- that these random regions can themselves include rare regions of anomalously low disorder which are approximately homogeneous and consequently favour thermalisation, however I will not consider the effect of these regions in detail here. Note that while the homogeneous regions would in isolation be integrable, as a consequence of the disorder-free XXZ model also being integrable, this is not expected to have a significant effect except in the limit of $p \to 0$, similarly to the $d \to 0$ limit of the conventionally disordered XXZ chain.

The second case, which I shall call \emph{dilute binary disorder}, is similarly dependent on a probability $p$ but in this case can take only two values, denoted $W_0$ and $d$:
\begin{align}
h _i =
    \begin{cases}
      W_0, & \text{with probability}\ (1-p), \\
      d, & \text{with probability}\ p.
    \end{cases}
\end{align}
For $p=0.5$, this reduces to a straightforward bimodal distribution (similar to others studied in the context of MBL~\cite{Kshetrimayum+19}), while in the limits $p \to 0$ and $p \to 1$ the model is translationally invariant and integrable. We can anticipate that the choice $p=0.5$ will be the most likely value to host a stable localised phase, however we shall see later that in fact the hallmarks of localisation exist over a wider region of parameter space than one might expect. In the following I set $W_0=1$ throughout, and vary $d$.

The effects of dilute regions have been studied previously in the context of MBL in the case of both rare strongly disordered impurities~\cite{Sels+21b} and small thermalising regions~\cite{Goihl+19,Luitz+17} or baths~\cite{Sels22,Morningstar+22}, however this work takes a somewhat different approach. Rather than studying the effect of a single anomalous region on the rest of the spin chain, as in previous works, here I look at the effect of changing the density of such thermalising regions in a way that aims to mimic Griffiths-type effects close to the localisation transition. In particular, here the transition will be driven by tuning the distribution of thermalising regions, rather than by tuning the bandwidth of the disorder distribution (i.e. disorder strength) as is more conventionally done. This provides a controllable way of introducing the resonant regions which are critical for the many-body delocalisation transition~\cite{Thiery+18}.

\section{Methods}

I will employ full exact diagonalisation using the \texttt{QuSpin} package~\cite{Weinberg+17,Weinberg+19}, and will examine both the non-equilibrium dynamics and the level spacing statistics. Both are considered standard measures of localisation effects in many-body quantum systems. 
In all of the following, the exact diagonalisation results have been averaged over a minimum of $N_s = 512$ disorder realisations (with up to $N_s=4096$ for the smallest system sizes used in this work).
    
The first quantity of interest is the disorder-averaged level-spacing ratio, which is by now a standard measure of investigating localisation and chaos in quantum systems~\cite{Oganesyan+07}. It is defined as:
\begin{align}
\delta_{\alpha} &= | \varepsilon_{\alpha} - \varepsilon_{{\alpha}+1} |, \\
r_{\alpha} &= \textrm{min}(\delta_{\alpha},\delta_{{\alpha}+1})/\textrm{max}(\delta_{\alpha},\delta_{{\alpha}+1}).
\end{align}
where the $\varepsilon_{\alpha}$ are the energy eigenvalues and $\delta_{\alpha}$ is the difference between successive energy levels. This quantity takes the value $r \approx 0.53$ in a delocalised phase (following Wigner-Dyson statistics, indicative of level repulsion), and $r \approx 0.39$ in a localised phase (following Poisson statistics, indicative of a random distribution of energy levels).

\begin{figure}[t!]
    \centering
    \includegraphics[width=\linewidth]{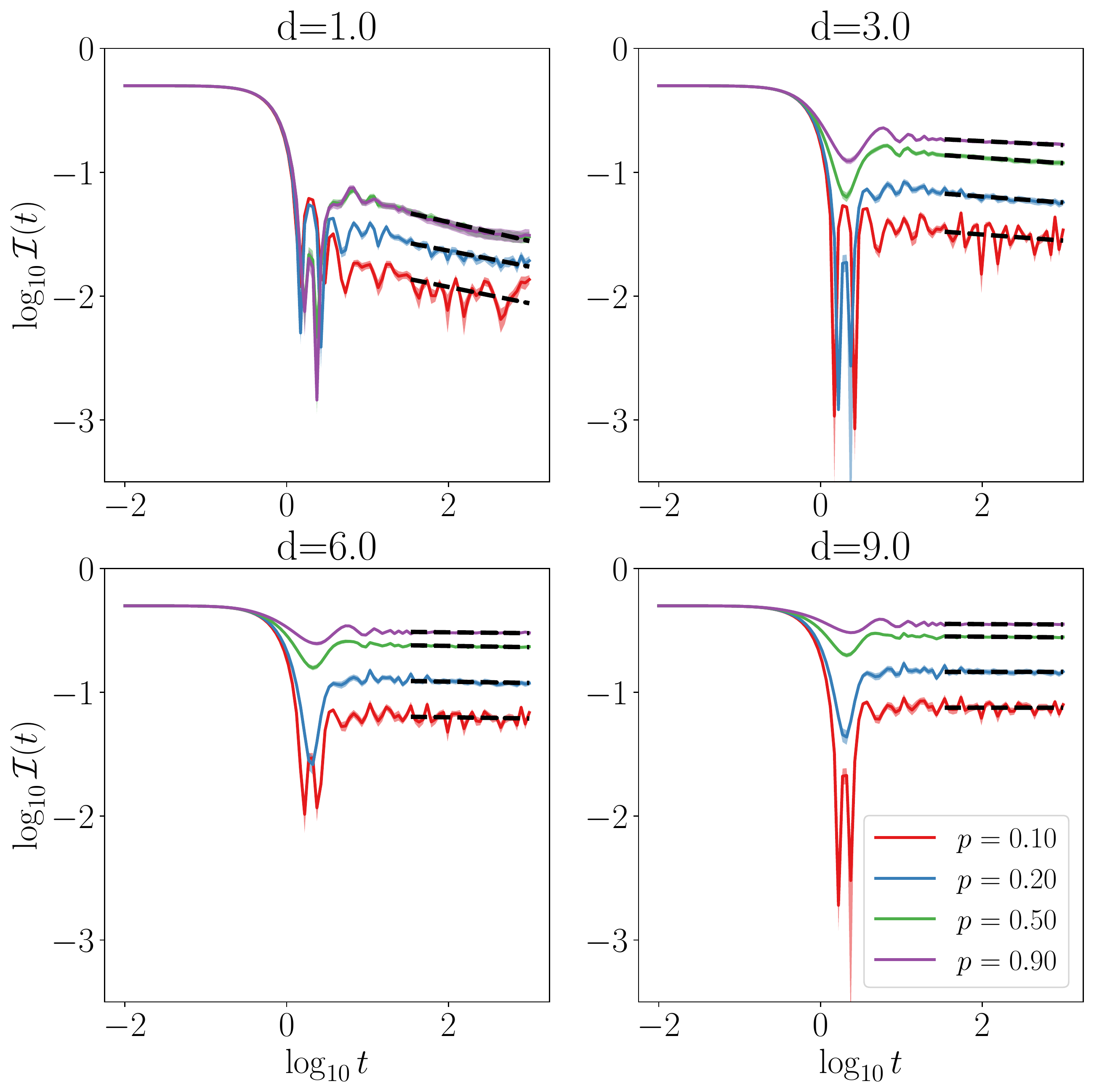}
    \caption{The decay of the imbalance (staggered magnetisation) following a quench from an initial Néel state, for $L=16$ and averaged over $N_s=1024$ disorder realisations. The four panels show four different disorder strengths, and the coloured lines each show different values of $p = 0.1,0.2,0.5,0.9$. The black dashed lines are power-law fits to the long-time behaviour, shown here as guides to the eye. The translucent shaded area around each line represents the uncertainty: note that in many cases it is roughly the same as the linewidth. (The data for $p=0.1$ with $d=1.0$ has been smoothed by convolution with a Gaussian filter of width $1\sigma$.)}
    \label{fig.ed_imb}
\end{figure}

I will investigate the non-equilibrium dynamics of the imbalance following a quench from a Néel state of the form $\ket{1010...}$, where $\ket{1} = \ket{\uparrow}$ and $\ket{0} = \ket{\downarrow}$. The imbalance (staggered magnetisation) is defined as:
\begin{align}
    \mathcal{I}(t) = \frac{1}{L} \sum_i (-1)^i \langle S^{z}_i(t) \rangle.
    \label{eq.imb}
\end{align}
such that $\mathcal{I}(0)=1/2$ and it decays in time. This observable gives us an idea of how much `memory' of its initial state the system has.
In order to further investigate the character of the transport on the delocalised side of the transition, I will also compute the infinite-temperature dynamical correlation function:
\begin{align}
    \mathcal{C}_i(t) &= 4 \langle S_i^z(t) S_i^z(0) \rangle.
    \label{eq.corr}
\end{align}
The thermal expectation value of an observable $O$ is defined as $\langle O \rangle = \textrm{Tr}[\exp(-\beta \mathcal{H}) O]/\textrm{Tr}[\exp(-\beta \mathcal{H})]$, where $\beta =1/T$ is the inverse temperature. In the limit of infinite temperature, this becomes $\langle O \rangle = \textrm{Tr}[O]/D$, where $D$ is the Hilbert space dimension. It has been shown, however, that rather than performing the trace over all basis states, we can make use of \emph{dynamical quantum typicality}~\cite{Goldstein+06,Reimann07,Bartsch+09,Bartsch+11,Elsayed+13,Richter+19,Heitmann+20,Chiaracane+21} to replace this with an expectation value with respect to a single randomly-chosen pure state of the form $\ket{\psi} = C \sum_k (a_k + i b_k) \ket{\phi_k}$, where $C$ is a normalisation constant, $a_k$ and $b_k$ are chosen randomly from Gaussian distributions of mean zero~\footnote{The distributions used in this work have a variance of $1/2$.} and the $\ket{\phi_k}$ are the basis states. This state can be considered a `typical' state of the desired ensemble, and the properties of this state representative of the full trace over the entire Hilbert space. The resulting expectation value with respect to this `typical' state then becomes:
\begin{align}
    \mathcal{C}_i(t) &= 4 \braket{\psi | S_i^z(t) S_i^z(0) | \psi} + \epsilon
\end{align}
where the final term is an error which has zero mean and a standard deviation that scales as $\propto 1/\sqrt{D}$ where $D$ is the Hilbert space dimension. For many-body systems, $D$ is exponentially large in the system size, guaranteeing that statistical fluctuations vanish rapidly as the system size increases. A different random state is chosen for each disorder realisation.
In the following, I shall set $\mathcal{C}(t) \equiv \mathcal{C}_{L/2}(t)$, such that boundary effects are minimised. 

In a phase with diffusive transport, one would expect to see this correlation function decay like $C(t) \sim 1/\sqrt{t}$, whereas if the transport is sub-diffusive it will instead decay like $\mathcal{C}(t) \sim t^{-\alpha}$ with $\alpha < 1/2$. 
It has been shown in Refs.~\cite{Agrawal+15,Khait+16} that at infinite temperature the exponent $\alpha$ is linked to the behaviour of the optical conductivity $\sigma(\omega) \sim \omega^{\beta}$ via the relation $\beta+2\alpha = 1$, therefore knowledge of this correlation function also gives us information about the optical conductivity. It was also demonstrated in Ref.~\cite{Agrawal+15} that it is extremely challenging to reliably extract an exponent $\alpha=1/2$ in the diffusive regime due to significant finite-size effects, and that extrapolation to the $L \to \infty$ limit is required in order to recover the expected value of $\alpha=1/2$. I shall make use of a similar procedure below.

\begin{figure}[t!]
    \centering
    \includegraphics[width=\linewidth]{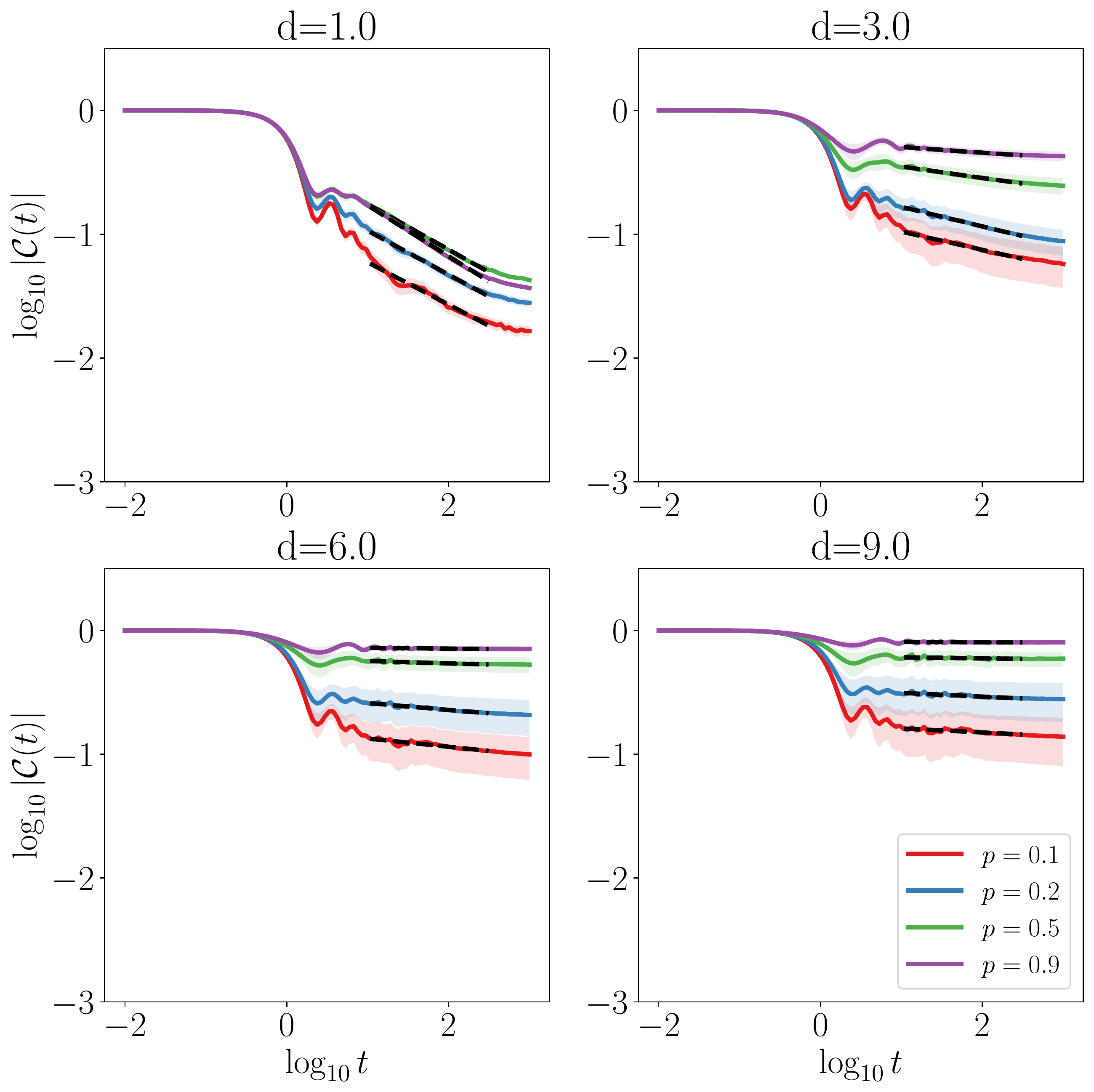}
    \caption{The decay of the infinite-temperature correlation function for $L=16$ with $N_s=512$, again for four different disorder strengths and varying values of the dilution $p$. As before the shaded area represents the uncertainty, and the dashed lines are power-law fits shown as guides to the eye.}
    \label{fig.ed_corr}
\end{figure}

I will also compute the bipartite von Neumann entanglement entropy density $\mathcal{S}(t)$ across a cut in the centre of the chain:
\begin{align}
    \mathcal{S}(t) = -\frac{2}{L} \textrm{Tr}_{\mathcal{A}}[\rho_A(t) \log(\rho_{A}(t))].
\end{align}
where $\rho_{A}(t) = \textrm{Tr}_{\mathcal{B}}[\ket{\psi(t)}\bra{\psi(t)}]$ is the reduced density matrix of one half of the chain (subsystem $\mathcal{A}$, with length $L/2$) after tracing out the remaining sites in subsystem $\mathcal{B}$. The entanglement entropy has previously been shown to increase logarithmically in time in the MBL phase~\cite{Bardarson+12,Kim+14}, in contrast to the much faster growth expected in an ergodic system, and provides an important indicator of the slow dynamics characteristic of MBL. In particular, unlike local observables such as a persistent density imbalance, the entanglement entropy is capable of clearly distinguishing true many-body localisation from Anderson localisation.

\section{Results}

\begin{figure}
    \centering
    \includegraphics[width=\linewidth]{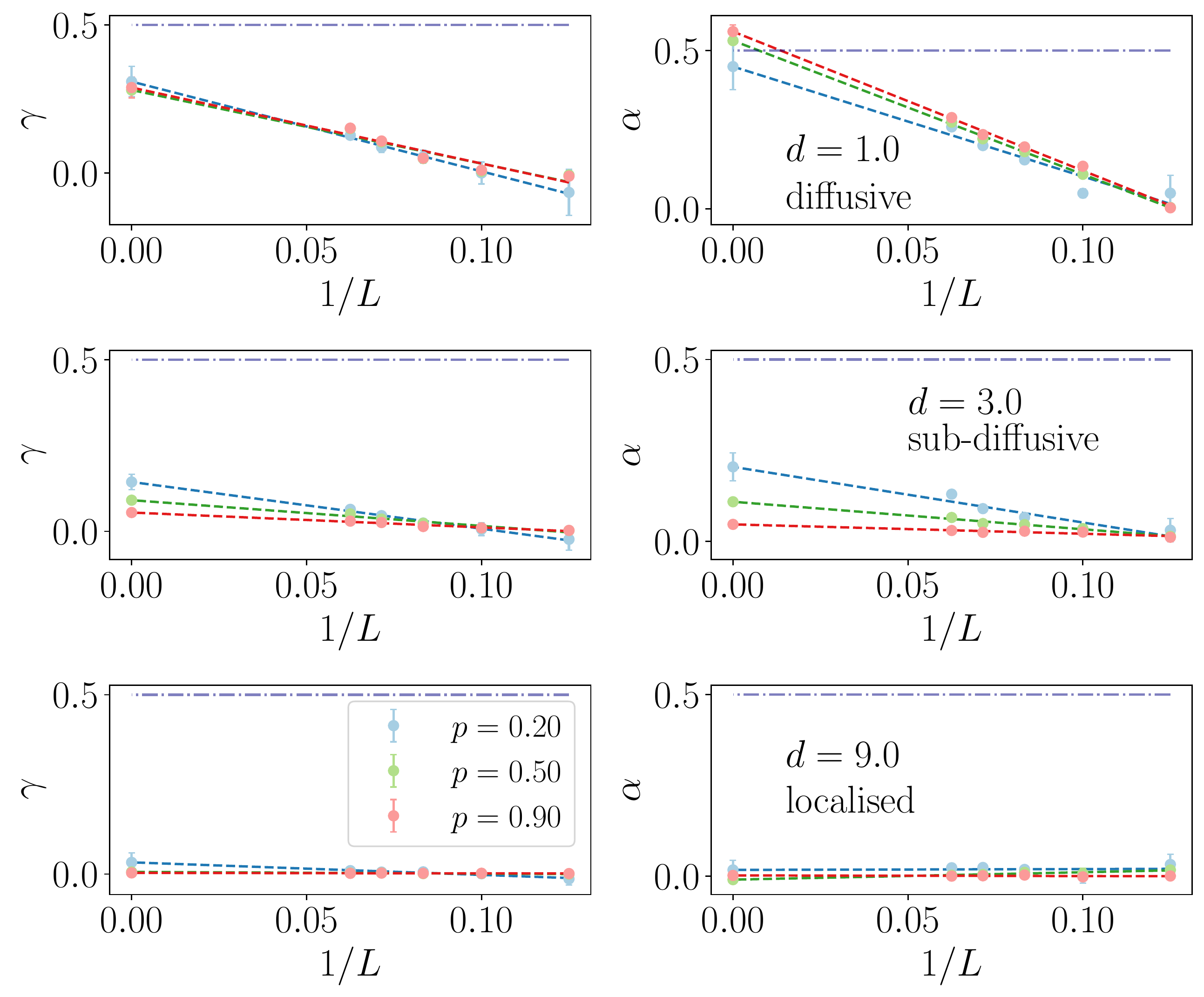}
    \caption{Decay exponents from fits to the imbalance $\mathcal{I}(t) \propto t^{-\gamma}$ and infinite-temperature correlation function $C(t) \propto t^{-\alpha}$ respectively, shown for three different disorder strengths ($d=1.0,3.0,9.0$, top to bottom) and three different values of the dilution parameter $p$. The dashed lines show the linear extrapolation to the infinite system size limit, where for diffusive transport one would expect to find $\alpha = 1/2$, indicated by the purple dot-dashed line. For weak disorder, the results are consistent with diffusive transport, but as the disorder is increased, the decay becomes significantly slower with $\alpha, \gamma \to 0$, suggestive of non-diffusive transport at intermediate disorder and full localisation at strong disorder. Error bars reflect the uncertainty in the fits, and are in many cases smaller than the plot markers.}
    \label{fig.dilute_random_exponent}
\end{figure}

\subsection{Dilute Random Disorder}
A summary of the results for dilute random disorder are shown in the phase diagrams of Fig.~\ref{fig.dilute_random}. Firstly, the behaviour in both limits $p \to 0$ and $p \to 1$ is as expected, yielding a vanishing imbalance in the former case (characteristic of the diffusive transport expected in the integrable XXZ model~\cite{Mukerjee+06}) and a localisation transition in the latter case, located at approximately $d_c \approx 3.7$~\cite{Luitz+15}. As this transition has been widely studied elsewhere, I do not linger on it, nor attempt to locate it any more precisely than has already been done in other works. For our purposes, it is sufficient to note that there is a localisation transition and ask what happens to it as $p$ is changed.

The most interesting physics occurs at disorder strengths $d>d_c$ when $p$ is reduced from unity, and one can examine what happens to the localisation transition. Fig.~\ref{fig.dilute_random}a) shows that the transition shifts to larger values of $d$ as more homogeneous regions are added, which makes intuitive sense as these regions act to favour thermalisation. The smaller $p$ becomes, the larger these thermalising regions are and the stronger their effects. Remarkably, however, signatures of a finite imbalance persist even to small values of $p \approx 0.2$. This effect is even more starkly revealed in the level spacing statistics, shown in Fig.~\ref{fig.dilute_random}b), where at strong disorder there are no areas of the phase diagram where $r$ takes on the Wigner-Dyson value of $\sim 0.53$ which would be characteristic of a chaotic ergodic phase. This this may be due to rare but strong impurities acting as bottlenecks, or simply a consequence of the finite resolution of the data shown in Fig.~\ref{fig.dilute_random}.

More information about how the system approaches its long-time state can be extracted from the evolution of these observables, rather than just their long-time limits. Fig.~\ref{fig.ed_imb} shows the evolution of the imbalance for a variety of different disorder strengths and choices of the dilution parameter $p$, on a double-log scale with power-law fits to the late-time behaviour indicated by black dashed lines. In all figures, the uncertainty shown for each observable $\mathcal{O}$ is obtained by computing the mean $\overline{\mathcal{O}}$ and the variance $\sigma^2(\mathcal{O})$, and plotting the range given by $\overline{\mathcal{O}} \pm \sigma^2(\mathcal{O})$ as a translucent shaded area: in the present figure, the uncertainty is close to the linewidth.

At low disorder (i.e. in the delocalised phase), the imbalance decays like a power-law at long times. This persists to intermediate disorder ($d=3.0$) although the slope of the power-law is significantly reduced, implying a slower decay. At the two highest disorder strengths shown in Fig.~\ref{fig.ed_imb}, the imbalance appears to saturate at late times and does not exhibit power-law decay, although it must be noted that for small values of $p$ the imbalance does indeed saturate close to zero, which is inconsistent with a fully many-body localised phase.

\begin{figure}
    \centering
    \includegraphics[width=\linewidth]{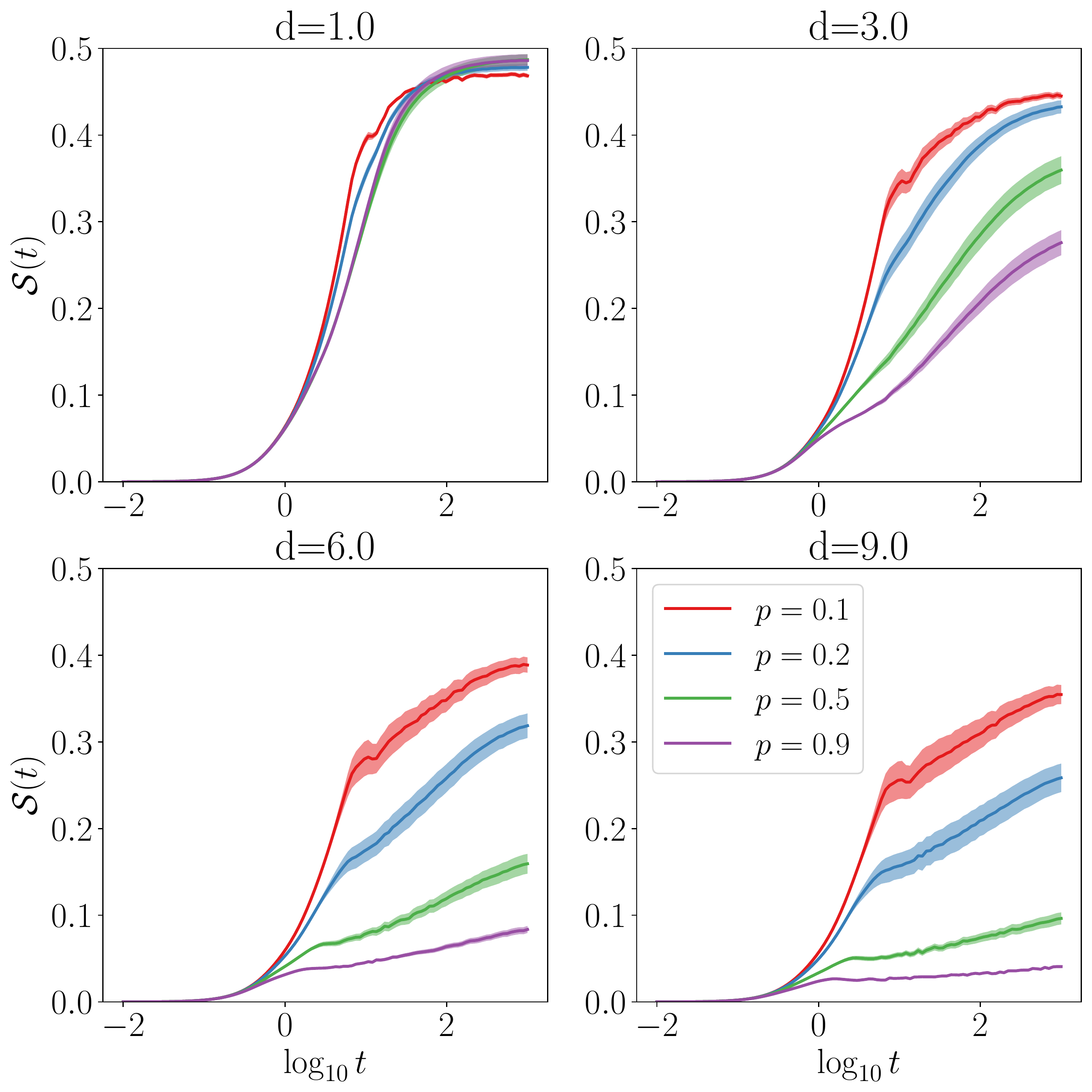}
    \caption{Growth of the entanglement entropy following a quench from an initial Néel state, again for four different disorder strengths and varying values of dilution $p$, computed for $L=16$ using exact diagonalisation. The shaded area around each curve again represents the uncertainty (variance). For the strongest disorder strength ($d=9.0$), even at the weakest value of the dilution ($p=0.1$) there is still a slow logarithmic growth of the entanglement entropy at late times, suggesting that the rare impurities act as significant bottlenecks to the growth of entanglement.}
    \label{fig.ed_ent}
\end{figure}

\begin{figure*}[t!]
    \centering
    \includegraphics[width=\linewidth]{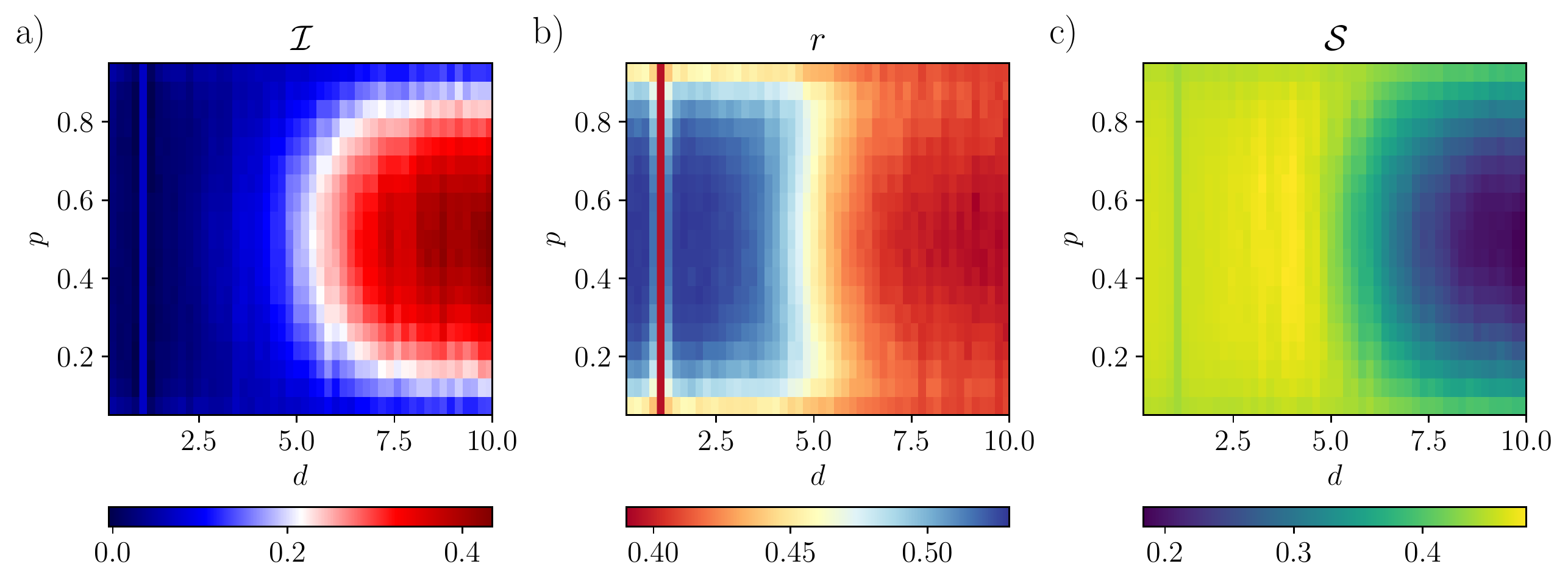}
    \caption{Numerical results for the case of dilute binary disorder, for a system size $L=14$ and averaged over $N_s=512$ disorder realisations. a) The imbalance long after a quench from a Néel state ($tJ_0=10^3$). b) The averaged level spacing statistics, which vary from $r \sim 0.39$ in a localised phase to $r \sim 0.53$ in an ergodic phase. c) The entanglement entropy a long time after the quench. Note the singular behaviour along the integrable line $W_0=d=1$, clearly visible in all three panels as a vertical line.}
    \label{fig.dilute_binary}
\end{figure*}

\begin{figure}
    \centering
    \includegraphics[width=\linewidth]{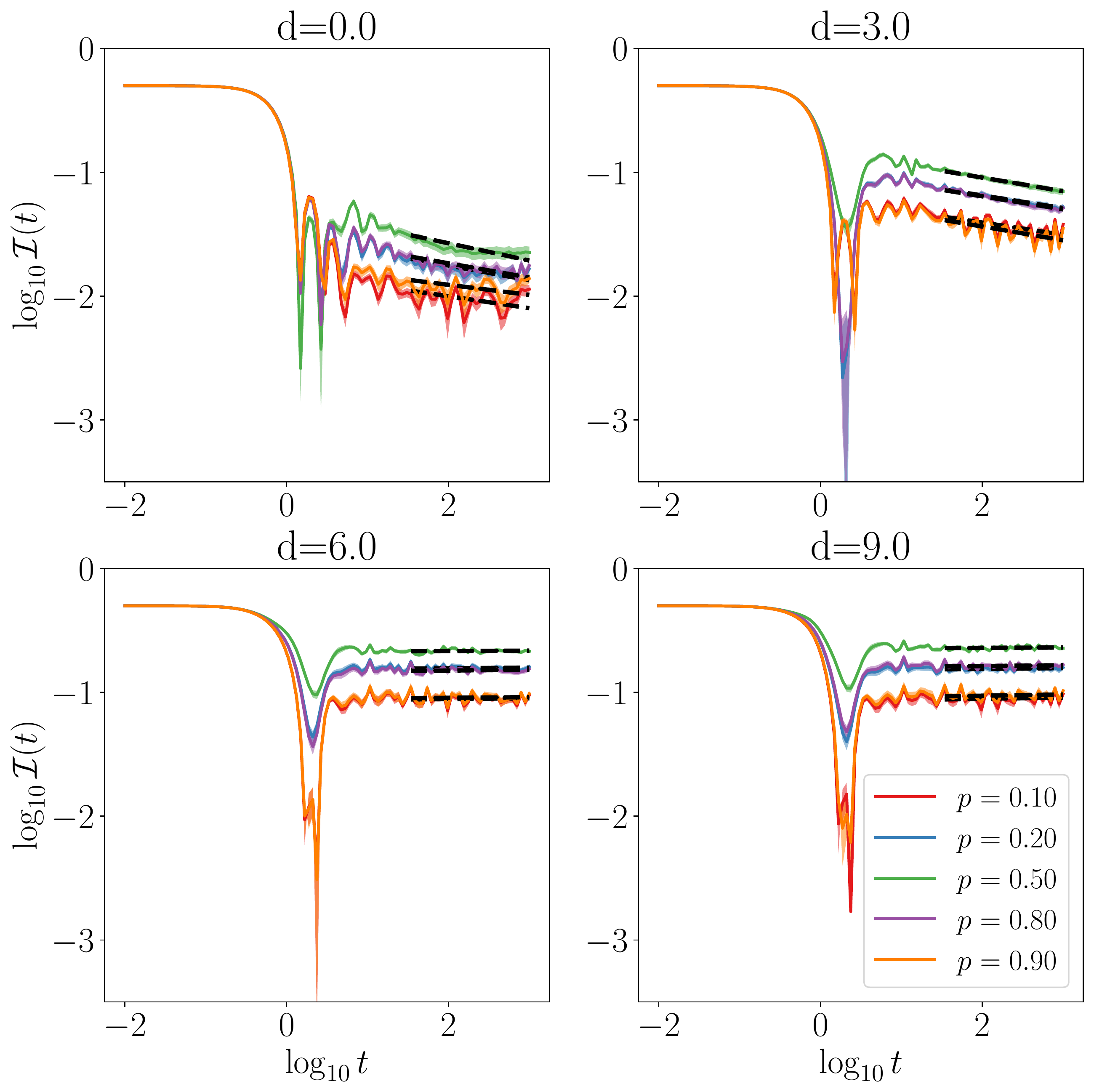}
    \caption{The decay of the imbalance (staggered magnetisation) following a quench from an initial Néel state in the case of dilute binary disorder, for $L=16$ and $N_s=1024$. Note that the $p=0.2$ and $p=0.8$ lines lie almost on top of each other due to the symmetry of the phase diagram, as do the lines for $p=0.1$ and $p=0.9$. (For $d=0.0$, the $p=0.1$ and $p=0.9$ data have been smoothed by convolution with a Gaussian filter.)}
    \label{fig.ed_imb2}
\end{figure}

\begin{figure}
    \centering
    \includegraphics[width=\linewidth]{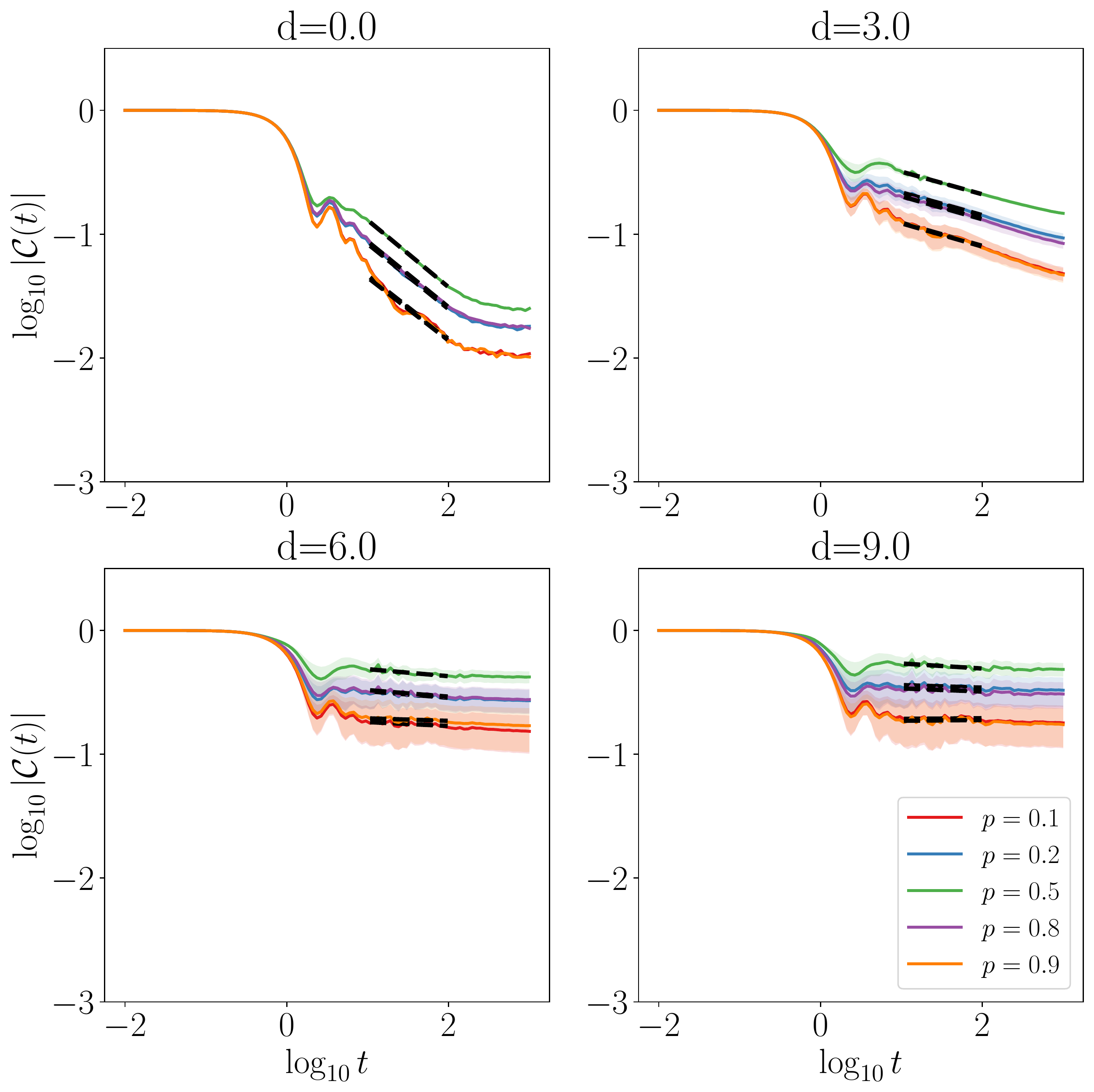}
    \caption{The decay of the correlation function following a quench from an initial Néel state in the case of dilute binary disorder, for $L=16$ with $N_s=1024$, again for four different disorder strengths and varying values of the dilution $p$. As before the shaded area represents the uncertainty. Again, the $p=0.2$ and $p=0.8$ lines overlap due to the symmetry of the phase diagram, as do the lines for $p=0.1$ and $p=0.9$.}
    \label{fig.ed_corr2}
\end{figure}

This behaviour can be corroborated via the correlation function, shown in Fig.~\ref{fig.ed_corr}, which exhibits qualitatively similar behaviour. Following Ref.~\cite{Agrawal+15}, here the fits are performed over an intermediate window in time before the correlation function saturates to its long-time value. Again, there is a power-law decay at low disorder but a clear plateau at large disorder strengths, even for small values of the dilution parameter $p$, indicating that the long-time behaviour of the system is strongly affected by the presence of these rare yet large bottlenecks. 

It is possible to extract decay exponents from the imbalance $\mathcal{I}(t) \propto t^{-\gamma}$ and correlation function $C(t) \propto t^{-\alpha}$ respectively, as shown in Fig.~\ref{fig.dilute_random_exponent}. While the decay of the imbalance is not directly related to the diffusion exponent $\alpha$, it is nonetheless instructive as it displays the same qualitative behaviour. In some cases, it has even shown to exhibit quantitatively similar behaviour. The decay exponent for the correlation function approaches the diffusive value of $\alpha = 1/2$ in the thermodynamic limit, as expected, however convergence as a function of system size is extremely slow and extracting reliable exponents is challenging. Nonetheless, the results indicate that for weak disorder ($d=1.0$), the decay of the correlation function is consistent with diffusive behaviour, as expected. In all cases, the imbalance exhibits a smaller exponent $\gamma < \alpha$, perhaps due to finite-size effects or the specific choice of initial state. For intermediate disorder ($d=3.0$) the decay is significantly and visibly slower, consistent with subdiffusive transport. It is also interesting to note that the exponent is suppressed at large values of $p$, vanishing almost entirely for $p=0.9$, and becoming much larger as the density of homogeneous regions is increased, confirming that the dilution parameter $p$ strongly changes the character of the transport in this model. At the largest disorder strength shown in Fig.~\ref{fig.dilute_random_exponent} ($d=9.0$), the exponents are both effectively zero for all values of $p$, suggestive of localisation, although it is not possible to rule out the possibility of extremely slow transport that cannot be resolved at these system sizes and evolution times~\cite{Panda+20}. In fact, it is very likely that this is the case, based upon the high density of thermalising regions which would by design overwhelm conventional perturbative arguments for the stability of MBL~\cite{Imbrie16a}. It should be noted, however, that the local integrability of the inclusions in this model could play a role in allowing signatures of localisation to persist to long times.

In addition, Fig.~\ref{fig.ed_ent} shows the growth of the half-chain entanglement entropy $\mathcal{S}(t)$ following the quench, again for a variety of disorder strengths and values of the dilution parameter. At weak disorder, the entanglement entropy grows rapidly and quickly saturates for all values of the dilution parameter $p$, while at intermediate disorder the dynamics separates into an initial transient period of fast growth followed by a slower logarithmic increase that persists to late times. Notably, this feature is also present at strong disorder and small values of dilution, for example $d=9.0$ and $p=0.2$, where the impurities are strong yet rare. This confirms again that the presence of even very rare disordered regions is enough to lead to the emergence of standard MBL phenomenology which predicts $\mathcal{S}(t) \sim \log(t)$~\cite{Bardarson+12,Kim+14}. The long-time values of $\mathcal{S}(t=10^3)$ are shown in Fig.~\ref{fig.dilute_random}c), where it is clear that the entanglement in the localised region is significantly smaller than elsewhere in the phase diagram. 
The fast linear growth of the entanglement entropy at short times results from the nearly-free transport of spin over a region set by the localisation length (for $p \to 1$) or the size of the homogeneous regions $(p \to 0$), and consequently grows more for small values of the dilution parameter $p$ where spin transport over larger distances is possible. At later times, however, the disordered regions always appear to dominate the dynamics even for small values of $p$, leading to the slow logarithmic growth characteristic of MBL.

\subsection{Dilute Binary Disorder}
I now contrast the results shown in the previous section with the corresponding results for dilute binary disorder, which are summarised in Fig.~\ref{fig.dilute_binary}. In the limits $p \to 0$ and $p \to 1$ the system behaves in an integrable manner, with vanishing imbalance and a level spacing parameter consistent with Poisson level statistics, as expected for an integrable system. At the maximally disordered value of $p=0.5$, there is a transition or crossover between an ergodic and a localised phase at a critical $d_c \approx 4$, similarly to the $p \to 1$ limit of the model with dilute random disorder. Moving away from this point, as $\delta p = |0.5-p|$ increases from zero, the localised phase is quickly destabilised, however once again the level statistics do not significantly deviate from their Poisson value of $r \sim 0.39$ despite the imbalance vanishing, consistent with a crossover from a disordered to an integrable regime.

Note that along the line $W_0=d=1$, the system is integrable for all values of $p$ and displays a rapid transition to a chaotic ergodic phase as $\delta W = |W_0-d|$ becomes non-zero. This is not an error, and is simply a reflection that there is no `disorder' to speak of along this line, as the system is translationally invariant (and integrable) regardless of the value of $p$.

\begin{figure}
    \centering
    \includegraphics[width=\linewidth]{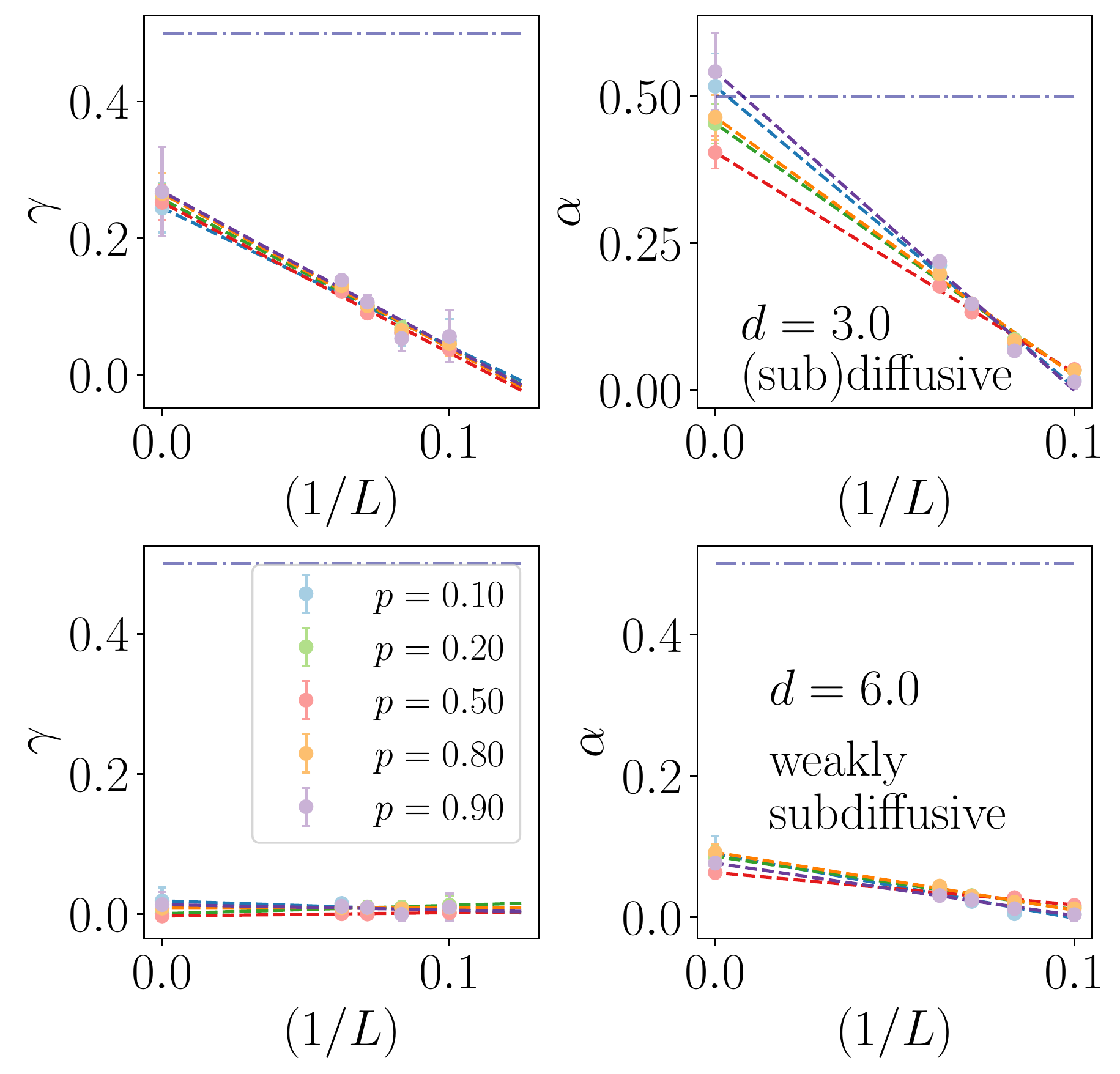}
    \caption{Decay exponents from fits to the imbalance $\mathcal{I}(t) \propto t^{-\gamma}$ and correlation function $C(t) \propto t^{-\alpha}$ respectively, shown for two different disorder strengths ($d=3.0$ and $d=6.0$) and three different values of the dilution parameter $p$. The dashed lines show the linear extrapolation to the infinite system size limit, with the diffusive value $\alpha = 1/2$ indicated by the purple dot-dashed line. For $d=3.0$ the results are consistent with diffusive transport in the limits of $p \to 0$ and $p \to 1$, with subdiffusive transport inbetween, with the smallest value of $\alpha$ occuring for $p=0.5$. For $d=6.0$, the imbalance suggests that transport is almost entirely absent, but the correlation function indicates that slow subdiffusive transport persists.}
    \label{fig.dilute_binary_exponent}
\end{figure}

\begin{figure}[t]
    \centering
    \includegraphics[width=\linewidth]{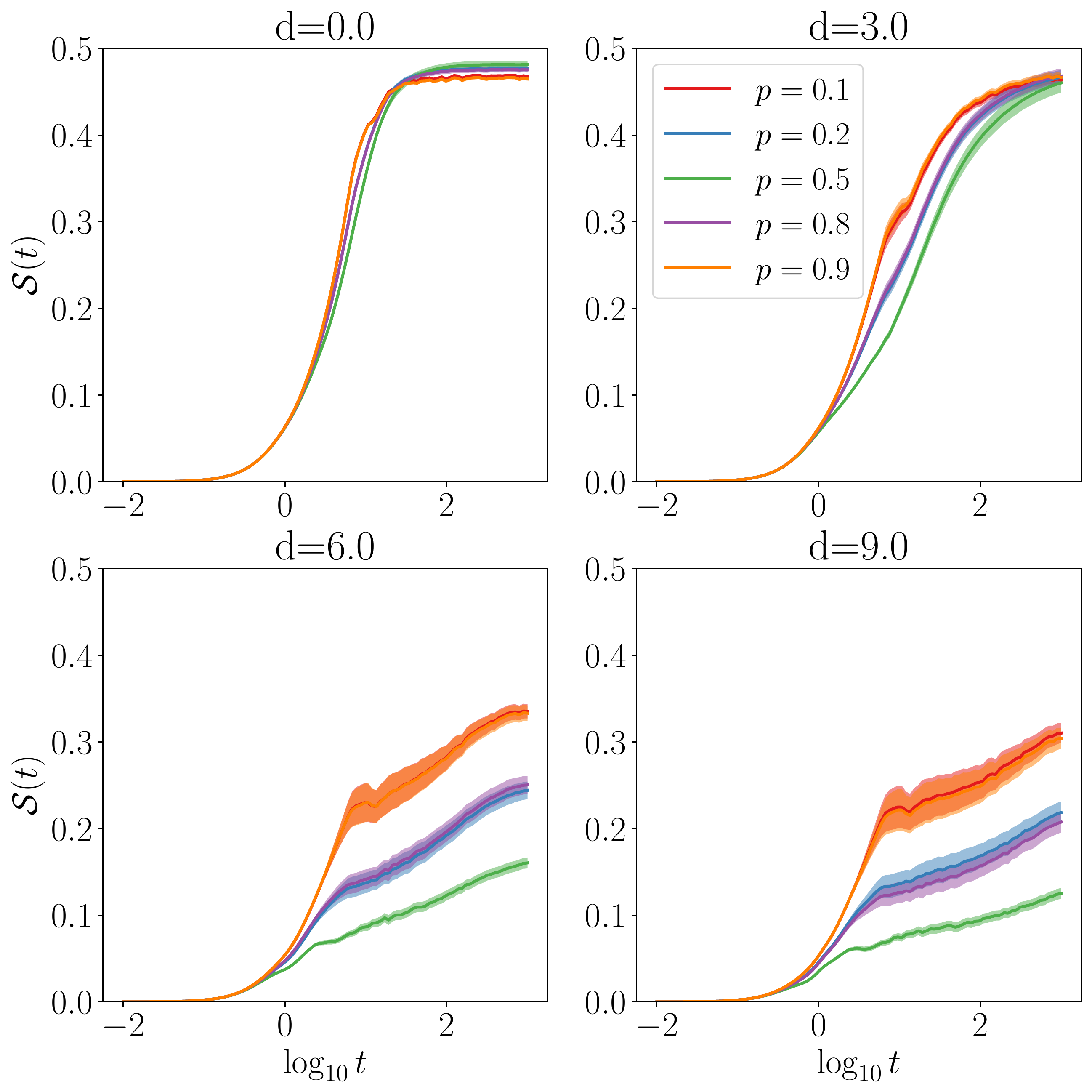}
    \caption{Growth of the entanglement entropy following a quench from an initial Néel state in the case of dilute binary disorder, again for four different disorder strengths and varying values of dilution $p$, computed for $L=16$ using exact diagonalisation. Note that the results for $p=0.1$ and $p=0.9$ are quantitatively very similar, as are the results for $p=0.2$ and $p=0.8$, mirroring the symmetry of the phase diagrams shown in Fig.~\ref{fig.dilute_binary}.}
    \label{fig.ed_ent2}
\end{figure}

As before, I compute the same set of observables, beginning with the imbalance following a quench from a Néel state, shown in Fig.~\ref{fig.ed_imb2}. The results are qualitatively similar to the case of dilute random disorder, showing a power-law decay of the imbalance at intermediate disorder and a flat `frozen' imbalance at large disorder strengths, suggestive of localisation. The value $d=1.0$ considered in the previous section is replaced here with $d=0.0$, as the system is integrable at $d=1.0$ for all values of $p$ (as $W_0 = d = 1.0$ and so the behaviour is qualitatively different here than for all other choices of $d$). Fig.~\ref{fig.ed_corr2} shows the decay of the dynamical correlation function, and again it exhibits familiar behaviour consistent with slow transport at intermediate disorder and localisation at higher disorder strengths. The extracted decay exponents are shown in Fig.~\ref{fig.dilute_binary_exponent}. At $d=3.0$, transport is subdiffusive for $\delta p \approx 0$, crossing over to diffusive for $\delta p \to 0.5$. At stronger disorder ($d=6.0$), transport is (weakly) subdiffisive for all values of $p$, where `weakly' in this context means that the exponent is close to zero. Again the transport is slowest for $p=0.5$, where the effects of disorder are most strongly felt. The effect of disorder is qualitatively similar as in the dilute random case, with strong disorder ultimately suppressing transport regardless of the value of the dilution $p$. Finally, Fig.~\ref{fig.ed_ent2} shows the entanglement entropy, where the behaviour is qualitatively consistent with the other observables, at large disorder strengths exhibiting a slow logarithmic growth around $p=0.5$, and a faster growth for $\delta p \to 0.5$ with late-time behaviour that appears to weakly saturate. Interestingly, even for values of $\delta p \approx 0.4$, the growth of the entanglement entropy remains slow for a long period of time, although the growth appears to undergo a weak crossover to faster-than-logarithmic at late times. This suggests that even very rare impurities can be significant bottlenecks to transport, provided they are large enough, and could indicate the formation of a metastable or prethermal phase which appears localised on intermediate timescales, before eventually thermalising.

\section{Discussion and Conclusion}

In this work I have studied two unconventional choices of disordered potentials and examined the stability of many-body localisation in both cases. The results indicate that localisation appears to be surprisingly robust towards the addition of homogeneous regions which should favour thermalisation, with signatures of localisation persisting to very long timescales. Even in parameter regimes where disordered regions are much smaller (on average) than homogeneous regions, localisation effects persist to the largest system sizes and longest times studied here. Based on analytical arguments regarding the stability of many-body localisation~\cite{Thiery+18}, it is very likely that the localisation phenomena observed over much of the phase diagrams shown in this work will not survive in the limit of large systems and very long times, however it is striking that in many ways, these regions appear just as stable as the conventionally studied localised regime. The indistinguishability of localisation in the dilute and fully random regimes (i.e. $p=1$ for dilute random disorder) suggests that extremely slow mechanisms for delocalisation cannot be captured by numerically exact simulations on such small system sizes.

These numerical simulations will serve as a useful roadmap for future studies examining further exotic properties of spin chains with dilute disorder. Here I have established the broad structure of the phase diagram and a few key properties, but there are many interesting open questions which dilute disorder may be able to shed some light on. In this work, I have focused on small system sizes and numerically exact methods in order to establish the broad properties of these disorder distributions in the regime most commonly studied by the many-body localisation community. Further studies using tensor network techniques to access larger system sizes may be able to contrast the finite-size scaling behaviour of dilute and fully random disorder distributions in order to see whether the behaviour in the thermodynamic limit continues to behave indistinguishably, or whether the localisation in the presence of dilute disorder will be destabilised more easily. While the latter scenario is more likely, the former would have interesting implications for the stability of MBL in the thermodynamic limit. This work has not looked in detail at the universal properties of dilute disorder, nor whether it strongly modifies any critical properties of the phase transition. Future works using techniques such as real-space renormalisation group may be able to establish whether the localisation transition triggered by changing $p$ at fixed $d$ is in the same universality class as the localisation transition studied in conventional random potentials. It would also be interesting to see if future analytical studies will be able to rigorously establish some of the numerical properties observed in this work, as to date few analytical works have considered dilute disorder. Being able to controllably add `rare' regions may offer an additional parameter for analytical works to study the destabilisation of localisation by carefully controlling the properties of these inclusions in a more systematic way than is possible in conventional random potentials.

While undoubtedly somewhat finely tuned and unlikely to exist in nature, the potentials introduced in this work offer a way to controllably add thermalising regions into an otherwise disordered sample and may offer a novel path forward for the study of Griffiths-type effects in disordered systems by allowing `rare regions' to be included in a more controlled manner than in purely random systems. They are also entirely within the realm of current generation experiments, particularly ultracold atomic gas platforms, where these tailored potentials can be realised, for example using a spatial light modulator or digital mirror device. This could also allow the effects of these potentials to be experimentally investigated in two dimensions, where the existence of MBL is not yet firmly established and the fragile nature of the putative localised phase should be much more vulnerable to the controlled addition of thermal inclusions as proposed in this work. It would be interesting to see if sub-diffusive transport also emerges in this situation, or if the presence of any finite concentration of thermal inclusions is sufficient to immediately induce conventional diffusive transport in greater than one spatial dimension. The disorder distributions studied in this work aimed to smoothly interpolate between disorder-free and fully random models, however as a consequence of the use of the `standard model' of MBL, the physics at low dilution is close to integrable. In future work, it would be a useful next step to investigate dilute disorder in a model which explicitly breaks integrability in the $p \to 0$ limit. Another possible avenue of investigation would be to study the local integrals of motion in systems with dilute disorder, and specifically their distribution, as slow transport has previously been linked to a broad distribution of local integrals of motion in a non-interacting system~\cite{Vidmar+21}. Given the key role that local integrals of motion play in MBL phenomenology, and their link with anomalous transport properties such as those studied here, it seems likely that the distribution of local integrals of motion in systems with dilute disorder may be a fruitful direction for further study.

\section*{Acknowledgements}

This project has received funding from the European Union’s Horizon 2020 research and innovation programme under the Marie Skłodowska-Curie grant agreement No.101031489 (Ergodicity Breaking in Quantum Matter). I gratefully acknowledge helpful discussions with C.~Bertoni, J.~Eisert, A.~Kshetrimayum and A.~Nietner, as well as J.~Richter for useful comments regarding the dynamical correlation function, and comments on the manuscript from L.~Vidmar and I.~Khaymovich.
All data and code are available at \cite{dilute_code,data}.

\begin{figure}
    \centering
    \includegraphics[width=\linewidth]{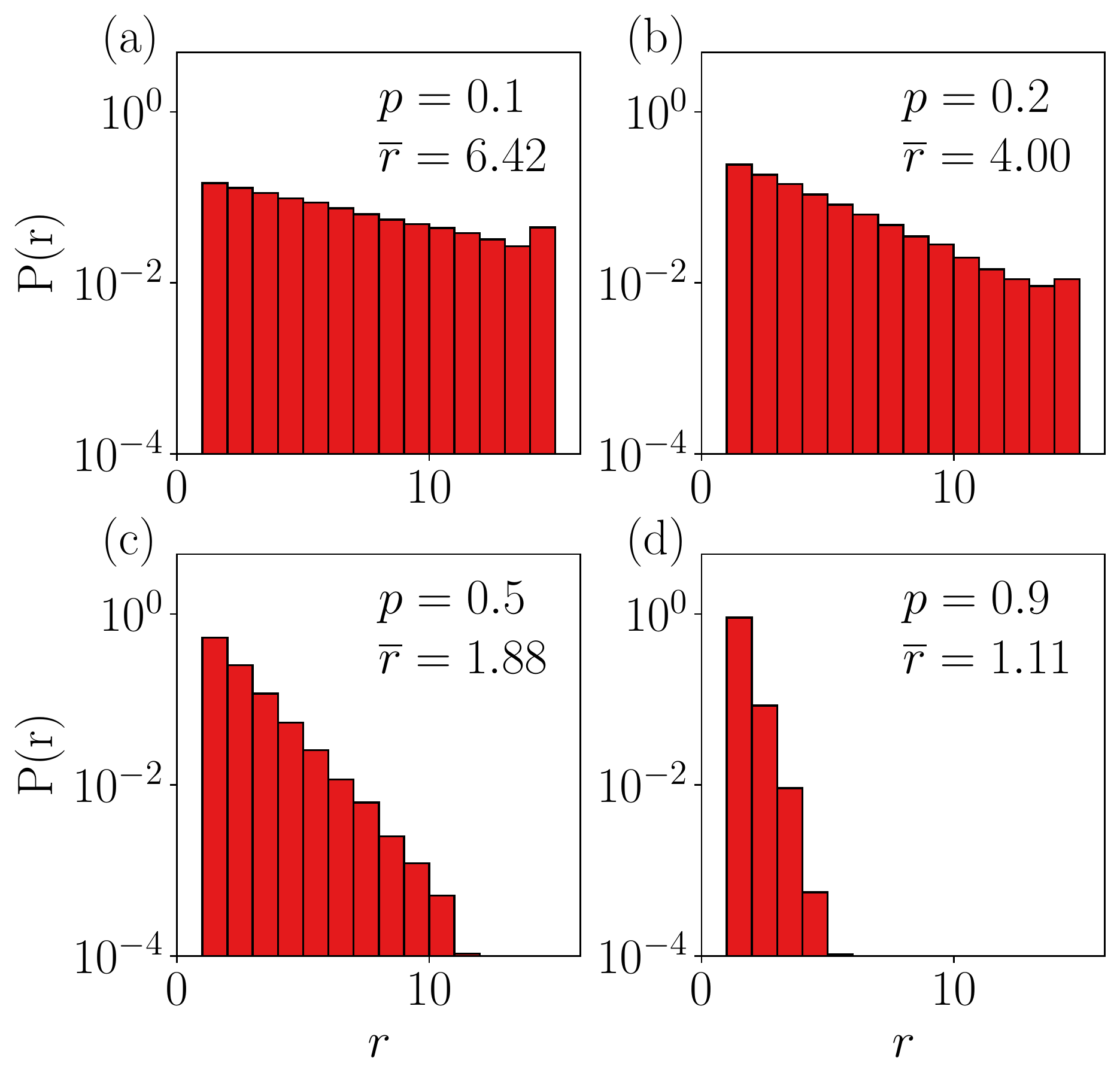}
    \caption{Normalised distributions $P(r)$ of the homogeneous (non-disordered) regions in the case of dilute random disorder, for various values of probability $p$, obtained for system size $L=16$ and averaged over $N_s=20,000$ randomly generated values of the potential. The mean length of homogeneous regions $\overline{r}$ is indicated in each panel. Note that the largest regions can be significantly larger than the mean value.}
    \label{fig.dist}
\end{figure}

\appendix

\section{Distribution of Rare Regions}
\label{app.dist}

The problem of analysing the distribution of rare regions can be mapped onto the equivalent problem of studying the results of coin tosses of a biased coin. For example, in the case of dilute random disorder, we can label the homogeneous lattice sites (which occur with probability $1-p$) as `heads', and the random sites (which occur with probability $p$) as `tails'. This has been extensively analysed in Ref.~\cite{Schilling90} and is a surprisingly rich problem despite its apparent simplicity.
Here we focus on the probability $P(r)$ to obtain a homogeneous region of length $r$. Numerical results are shown in Fig.~\ref{fig.dist}, for dilute random disorder with four different values of the probability $p$. The disorder strength is set to $d=9$, however its precise value is irrelevant here, as all we care about is whether the potential at a lattice site is random or otherwise (i.e. heads or tails). Here I average over $N_s=20,000$ randomly generated values of the potential for a system size $L=16$, consistent with the size used in the main text. (Note that simply computing the properties of the rare regions does not require fully diagonalising the Hamiltonian, only computing the on-site potentials, and therefore a very large number of samples can be used.) To a good approximation, the probability to obtain a homogeneous region of size $r$ typically decays exponentially, $P(r) \sim \exp(-r)$. The expected value for the largest homogeneous region is approximately given by $R_L(p) \approx \log_{1/p}(L(1-p))$, however the probability distribution for the size of the largest region (not shown) has a long tail, and in a given disorder realisation, the largest homogeneous region may be significantly larger~\cite{Schilling90}, therefore one must carefully consider the entire distribution of rare/homogeneous regions rather than only looking at averaged properties. A similar analysis can be conducted for dilute binary disorder, which I do not show here.

\bibliography{refs}

\end{document}